\documentclass[aps, a4paper,prl, twocolumn, amsmath]{revtex4}
\usepackage[T1]{fontenc}
\usepackage[utf8]{inputenc}
\usepackage[english]{babel}
\usepackage{amsmath}
\usepackage{amssymb}
\usepackage{graphicx}
\usepackage{caption}
\usepackage{lipsum}

\newcommand{\beq}{\begin{equation}}
\newcommand{\eeq}{\end{equation}}

\newcommand{\beqa}{\begin{eqnarray}}
\newcommand{\eeqa}{\end{eqnarray}}

\newcommand{\nn}{\nonumber}

\begin{document}
\title{\Large How should you discount your backtest PnL?}
\author{Adam Rej, Philip Seager \& Jean-Philippe Bouchaud}
\address{Capital Fund Management, 23 rue de l'Universit\'e, 75007 Paris}
\begin{abstract}
\it 
\center
{In-sample overfitting is a drawback of any backtest-based investment strategy. It is thus of paramount importance
to have an understanding of why and how the in-sample overfitting occurs. In this article we propose a simple framework that allows one to model and quantify in-sample PnL overfitting. This allows us
to compute the factor appropriate for discounting PnLs of in-sample investment strategies.}
\end{abstract}
\maketitle

\section{Introduction}
Data-driven systematic investment strategies are now widely employed by asset managers. Investment research teams sift through historical market data,
 in the hope of discovering recurring patterns that could be monetized. This approach to investing is appealing since rule-based decision-making process may
 be evaluated using statistical methods. Furthermore, it allows for overcoming common investing biases, e.g. loss aversion. 
 
\textit{Any} strategy, whether of discretionary or systematic nature, that is appraised using a backtest is at risk of being overfitted (see e.g. \cite{Harvey,Lopez,Paulsen}). In other words, part of its performance (or the entirety in extreme cases) is due to favorable alignment of market forces. This windfall performance is of course not to be counted upon in the future. In the language of stochastic processes, favorable or unfavorable market conditions simply represent pure noise. Taking favorable noise realizations at their face values is thus an important source of overfitting. There is however another equally important contributing factor. If the research team has reasons to believe that their strategy is sound, but the backtest P\&L does not meet their expectations, they will most likely not discard it willy-nilly. Instead, the strategy will be dissected into elementary building blocks and each one of them will be carefully studied. The research team will propose ``improvements'' to these building blocks and one or more series of improvements will result in an acceptable backtest performance. Did this procedure truly improve the strategy? In most cases the answer is no. The performance enhancement simply comes from ``improving'' the noise realization of the original strategy. Even if the improvement is genuine, there is no way of knowing. It is thus prudent to assume \textit{no improvement} as the base case scenario. In the following section, we shall concretize and quantify the above arguments.  

\section{The setup and main result} \label{sec:setup}

Let us assume that the researcher has identified a valid investment strategy. We shall consider this strategy in isolation, assuming
that it is de-correlated from any other known strategy. We will model the (log-)PnL of the strategy by a drifted Brownian motion
\beq \label{process}
{\rm d} \textrm{PnL} = \mu\, {\rm d}t + \sigma\,{\rm d}W 
\eeq
over a finite-time interval $(0,T)$. We measure $T$ in years. The true Sharpe ratio of the strategy $\textrm{SR}_t =\frac{\mu}{\sigma}$ remains unbeknownst 
to the researcher \cite{conv}. The best she can do is to calculate the estimated Sharpe ratio, $\textrm{SR}$. It is
well-known that the estimation of Sharpe ratios is subject to considerable errors because it is impossible to
separate the drift term in \eqref{process} from the realization of noise. The quality of the estimate may be increased
only by increasing the length of the backtest, $T$. In practice, however, for many asset types backtests are limited to (at most) couple 
of decades of daily data. Since for a $\textrm{SR}=0.5$ P\&L one needs 43 years of backtest data in order to be $99.9\%$ confident that the performance is significantly different than noise, it should be clear that statistical appraisal of lower Sharpe ratio strategies is fraught with risk. Some readers may be tempted to quip that there is no point in looking at such unattractive strategies to begin with. Recall, however, that here we consider investment strategies in \textit{isolation}. In practice, an asset manager would appraise the \textit{residual} performance with respect to existing strategies. Residual Sharpe ratios on the order of $0.3-0.5$ are commonplace in the CTA space.  

Before deploying a new strategy, the investment committee will examine several of its characteristics like risk-reward profile,
rebalancing frequency, maximum drawdown in backtest, correlation to other strategies, etc. A new strategy should increase
the diversification and the expected return of the portfolio. For a strategy uncorrelated with existing portoflio strategies, these requirements will translate into setting a Sharpe ratio threshold,
$\Theta$, that the strategy at hand needs to clear. Let us stress, however, that
the researcher is unable to distinguish between the process and its finite-time realization. This is one of the inherent
sources of overfitting, as she will only pitch strategies such that
\beq
\textrm{SR} > \Theta\,.
\eeq
We will thus assume that if the realization of the strategy \eqref{process} clears the threshold, it is presented
to the investment committee ``as is''. If the performance is below the required one, the researcher will try to improve
the strategy. These ``tweaks'' typically consist of slight modifications of the strategy, such as
replacing the filter with a similar one, changing some parameters, removing certain asset types, etc. The researcher
usually has a reasonably sounding narrative to justify these. Of course, only modifications
leading to improvement of the in-sample performance are retained. Here, we will assume that every such modification
deteriorates the out-of-sample performance. In other words, you cannot beat \eqref{process}. This may seem pessimistic,
but on average we believe it is not very far from reality. In any case, such a conservative assumption provides a well-defined
base case scenario. 

We now turn to modelling ``strategy improvements''. Let us divide the P\&L resulting from \eqref{process} into $N$ equal intervals. The Sharpe ratio
of the realization is the average Sharpe of its subsections
\beq \label{split}
\textrm{SR} = \frac{1}{N} \sum^N_{i=1} \textrm{SR}_i \,.
\eeq
In the case of discrete (daily) processes, this formula is a very good approximation as long as $\tfrac{T}{N}$ does not get too small. Sharpe ratios computed using a finite number of data points are approximately normally distributed  \cite{Lo}
\beq \label{section}
\textrm{SR}_i \in \mathcal{N}\left(\textrm{SR}_t, \sigma_{\textrm{SR}}(N) \right)\,,
\eeq
where
\beq \nn
\sigma^2_{\textrm{SR}}(N) = \tfrac{N}{T}\,\left(1 + \tfrac{\textrm{SR}^2_{t, \textrm{daily}}}{2}\right) \equiv N \sigma^2_{\textrm{SR}, \textrm{tot}}\,.
\eeq
Again, this approximation holds as long as there is a sufficient number of data points within each slice $\tfrac{T}{N}$. Observe that it is the daily version of $\textrm{SR}_t$ that enters the above formula and thus for all practical purposes the second term is negligible. 

Notice that the decomposition \eqref{split} may, but does not have to follow the chronological order. Thus for example, in the case of $N=5$ each slice may represent the performance on a given day of the week. We use index $i$ to label the slices, but this labelling does not imply any sort of ordering. 

\begin{figure}
	\includegraphics[scale=0.5]{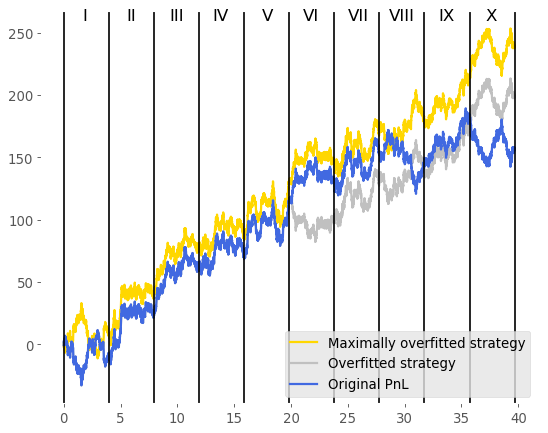}
	
	\caption{Here, the original P\&L is sliced in 10 sections, $f=0.3$. The maximally-overfitted trajectory is obtained by flipping signs in segments I, VIII and X. A non-maximal trajectory flips signs on a different subset of segments. } \label{fig:slicing}
\end{figure}

We shall assume that every modification (``tweak'') to the strategy translates into flipping predictor signs on a randomly chosen subset of ${f\times N}$ sections, see also \cite{moregen}. The parameter $0 \leq f \leq 1$ essentially captures the researcher's overfitting prowess. We illustrate this process in Figure \ref{fig:slicing}.

For a given realization of the original P\&L and a given binning into $N$ buckets, we call a modified strategy \textit{overfitted} iff its Sharpe ratio exceeds that of the original realization. If the set of overfitted strategies is non-empty, there always exists a ``maximally''-overfitted trajectory which maximizes the (in-sample) Sharpe ratio. The set of ``non-maximally''-overfitted trajectories consists of trajectories that are above the original P\&L and below the maximally-overfitted (MO) trajectory defined above. Of course, a large number of modified trajectories will in fact be \textit{worse} than the original P\&L. They will be discarded by the researcher. 

The Sharpe ratio of a modified strategy, $\textrm{SR}_{\textrm{m}}$ is given by
\beq \label{deco}
\textrm{SR}_{\textrm{m}} = -\frac{1}{N}\sum^{fN}_{i=1} \textrm{SR}_i + \frac{1}{N}\sum^{N}_{i= fN +1} \textrm{SR}_i\,.
\eeq

Since we assumed that the researcher only seeks to improve the strategy when its original realization underperforms, the quantity of interest is the probability density corresponding to
\beq \label{rho}
\textrm{p}\left(\textrm{SR}_{\textrm{m}} < x \,|\, \textrm{SR} < \Theta \right) = \int^{x}_{-\infty} \rho(y) dy\,.
\eeq

It is a straightforward to work out $\rho(y)$ using the decomposition \eqref{deco}. We present it in the Appendix.
Before we put $\rho(y)$ to work, we would like to gain some intuition about the parameters of our framework, $f$ and $N$. It is clear from
\eqref{deco} that the modified P\&L will be $(1-2f)$ correlated with the original strategy. The parameter $f$ thus captures how much the researcher is ready to depart from the original P\&L. We expect that typically, a researcher would not want the
``improved'' strategy to be less than 80\% correlated with the original proposal, which would translate into the upper bound of $f_{\textrm{max}} = 0.1$. Having said that, in what follows we shall keep $f$ as a parameter of the model. The parameter $N$, on the other hand,
does not have a simple, intuitive interpretation. It is thus very welcome news that the density $\rho(y)$, for Gaussian returns, does not depend on $N$. The reader may think of $N$ as a parameter that allows one to set up the scaffold, but that is no longer necessary once the construction is complete.  

We shall assume that if the original realization of the strategy does not clear the threshold, the research team will continue improving it and will stop as soon as $\textrm{SR}> \Theta$. While tampering with the parameters of the strategy is probably well modelled by flipping signs on a random subset of one P\&L binning, more radical interventions (for example changing filters from rolling averages to exponentially weighted moving averages) result in a new binning of the P\&L \eqref{split} followed by $f N$ sign flips on a randomly chosen subset of the new set of bins. As a consequence, there are no trajectories that one cannot tweak above the threshold $\Theta$. The average in-sample Sharpe ratio of the strategy presented to the investment committee is thus given by

\beqa \label{Einsample}
\mathbb{E} \left[\textrm{SR}_\textrm{in-sample}\right] &=&\,p(\textrm{SR} > \Theta ) \times \mathbb{E}_\mathcal{N} \left(\textrm{SR}\, |\, \textrm{SR} > \Theta \right)+ \nn \\
&&p(\textrm{SR} < \Theta)\times \mathbb{E}_{\rho} \left(\textrm{SR}_{\textrm{m}}\, |\, \textrm{SR}_{\textrm{m}} > \Theta \right). \nn \\
\eeqa
Both the probability $p$ and the conditional expectation value $\mathbb{E}_\mathcal{N}$ are computed under $\mathcal{N}\left(\textrm{SR}_t, \sigma_{\textrm{SR}, \textrm{tot}} \right)$. The conditional expectation $\mathbb{E}_{\rho}$ is calculated using $\rho$.
It is straightforward to define the expected out-of-sample Sharpe ratio. Tweaking the strategy \eqref{process} only degrades it
\beqa \label{Eoutofsample}
\mathbb{E} \left[\textrm{SR}_\textrm{out-of-sample}\right] &=&\,p(\textrm{SR} > \Theta)\times\textrm{SR}_t + \nn \\
&&p(\textrm{SR} < \Theta)\times (1-2f)\,\textrm{SR}_t.
\eeqa
We now may write down a closed-form formula for the overfitting factor (OFF)
\beq
\textrm{OFF} = \frac{\mathbb{E} \left[\textrm{SR}_\textrm{in-sample}\right]}{\mathbb{E} \left[\textrm{SR}_\textrm{out-of-sample}\right]}\,,
\eeq
which is the main result of this paper. It measures how much overfitting should be expected \textit{on average} if the researcher's behavior we assumed is representative. The overfitting factor exhibits dependence on $\textrm{SR}_t$, $\Theta$, $f$ and $T$.

\begin{figure}
	\includegraphics[scale=0.5]{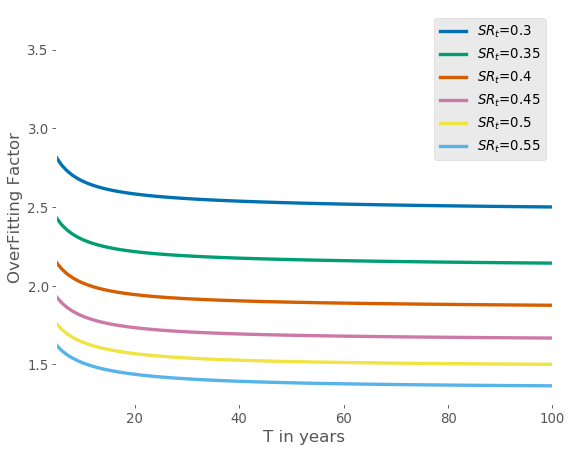}
	
	\caption{The overfitting factor as a function of $T$ and $\textrm{SR}_t<\Theta$. We fix the remaining parameters to $\Theta=0.7$ and $f=0.025$} \label{fig:SRs_vs_T}
\end{figure}

In Figure \ref{fig:SRs_vs_T} we explore the dependence on the length of the backtest. We observe that, for small values of $f$ at least, OFF diminishes in value for longer backtests and converges towards $\sim \tfrac{\Theta}{(1-2f)\textrm{SR}_t}$ for large values of $T$. This is because the probability that the original realization will cross the threshold drops with increasing $T$ and consequently the probability of the researcher's intervention increases accordingly, \textit{cf.} formulae \eqref{Einsample} - \eqref{Eoutofsample}. As $T$ becomes large, the conditional expectation $\mathbb{E}_{\rho} \left(\textrm{SR}_{\textrm{m}}\, |\, \textrm{SR}_{\textrm{m}} > \Theta \right)$ is then approximately $\Theta$.

\begin{figure}
	\includegraphics[scale=0.5]{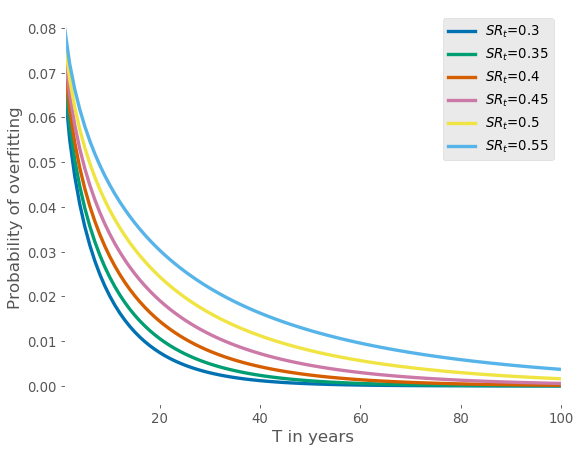}
	
	\caption{The probability of overfitting as a function of $T$ and $\textrm{SR}_t<\Theta$. We fix the remaining parameters to $\Theta=0.7$ and $f=0.025$} \label{fig:poof}
\end{figure}

Since effectively a fraction of the backtest is used for overfitting, increasing the length of the backtest also increases the overfitting freedom and the conditional expectation does not vary much. It is an entirely different story for probabilities. Both the probability that a one-off attempt will result in investment committee's acceptance
\beqa \label{PoA}
\textrm{PoA} &=&\,p(\textrm{SR} > \Theta) + \nn \\
&&p(\textrm{SR} < \Theta)\times p \left(\textrm{SR}_{\textrm{m}} > \Theta \right | \textrm{SR} < \Theta)\nn \\
\eeqa
and the probability of one-off overfitting, $\textrm{PoOF} = p \left(\textrm{SR}_{\textrm{m}} >\Theta \right | \textrm{SR} < \Theta)$, decrease with increasing $T$. We depict the latter in Figure \ref{fig:poof}. In practice, the longer the backtest the more reluctant the researcher should be to depart from the original P\&L. This would introduce an inverse relationship between $f$ and $T$ and would lead one to conclude that longer backtests decrease the level of overfitting. Observe, however, that such a relationship is behavioral and may not be derived from first principles.

\begin{figure}
	\includegraphics[scale=0.5]{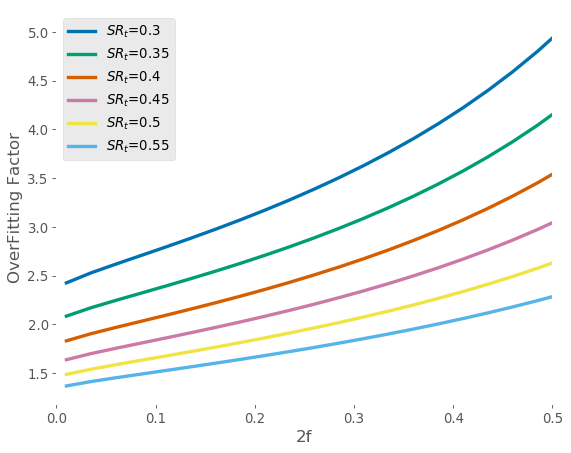}
	
	\caption{The overfitting factor as a function of $f$ and $\textrm{SR}_t<\Theta$. We fix the remaining parameters to $\Theta=0.7$ and $T=20y$} \label{fig:SRs_vs_f}
\end{figure}

In Figure \ref{fig:SRs_vs_f} we study how OFF varies as a function of the fraction $f$. As expected, the more overfitting freedom, the higher the level of overfitting. Lower Sharpe ratio strategies are more strongly impacted than the higher-Sharpe ones. The limit $f\to 0$ is tricky. Formulae \eqref{Einsample} - \eqref{Eoutofsample}  make sense iff $fN \geq 1$ and thus the limit $f \to 0$ may not be continuously reached. The probability of acceptance \eqref{PoA}, on the other hand, has a smooth $f \to 0$ limit because the probability of overfitting vanishes as $f \to 0$.

\begin{figure}
	\includegraphics[scale=0.5]{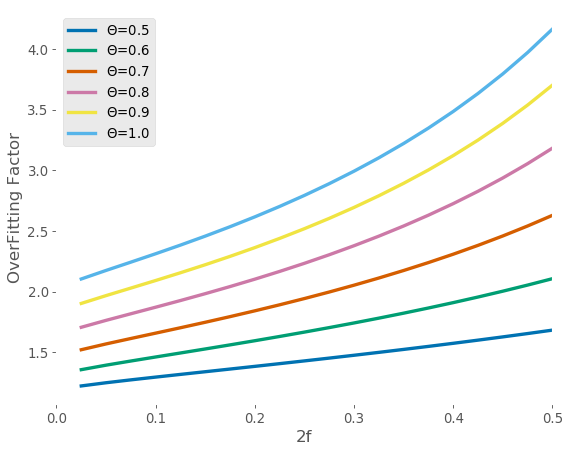}
	
	\caption{The overfitting factor as a function of $f$ and $\Theta$. We fix the remaining parameters to $\textrm{SR}_t=0.5$ and $T=20y$} \label{fig:SRt_vs_f}
\end{figure}

Finally, in Figure \ref{fig:SRt_vs_f} we plot the relationship between OFF and $f$ for different threshold Sharpe ratios but fixed $\textrm{SR}_t$ and the backtest length $T$. Setting the bar higher increases the level of overfitting, as expected.   
\section{Conclusions and outlook}
We have proposed an intuitive framework that offers a better insight into P\&L overfitting and how to quantify it. In particular, we have defined the overfitting factor which should be used to discount in-sample P\&Ls. We find that for typical Sharpe ratios of CTA strategies ($0.3-0.5$) and for reasonable values of other parameters ($f \sim 0.05$, $\Theta \sim 0.7$) the discounting factor is $\simeq 2$, which is in line with our experience and seems to be the industry standard \cite{Harvey}. Note, however, that the results are general and apply, in particular, to fast, high-Sharpe strategies. For such strategies, however, the threshold Sharpe ratio $\Theta$ is very sensitive to holding period (or gain per trade) of the strategy and is bounded from below by the breakeven Sharpe ratio. 

Finally, we would like to stress that we have made certain assumptions about researchers' \textit{modus operandi}. Our framework is however flexible enough to accommodate other overfitting patterns. For example, one could imagine that the researcher always maximally overfits. The expected in-sample Sharpe ratio would then be the average Sharpe ratio of the maximally overfitted trajectory. The dependence on $N$ would not go away in such case, as finer and finer tranching of the original P\&L improves the in-sample Sharpe ratio. 

\section{Acknowledgements}
We would like to thank Yann von Hansen, Adam Majewski, Emmanuel Serie and Gilles Z\'{e}rah for stimulating discussions. 
\appendix
\section{Appendix}
\section{The conditioned density}
The derivation of the density \eqref{rho} relies on the following observation. Since all returns are Gaussian, one may introduce two independent Gaussian random variables $U$ and $V$ defined as
\[
U := \frac{1}{N}\sum^{fN}_{i=1} \textrm{SR}_i, \qquad V:= \frac{1}{N}\sum^{N}_{i=fN +1} \textrm{SR}_i\,,
\]
with Gaussian pdf $P(U)$ and $Q(V)$. Using decomposition \eqref{deco} we can thus write
\beqa
&&\rho(y) = \frac{1}{\Phi\left(\Theta, \textrm{SR}_t, \sigma_{\textrm{SR}, \textrm{tot}}\right)} \int^{\infty}_{-\infty} \int^{\infty}_{-\infty} dU dV \nn \\
&&P(U)Q(V) \theta\left(\Theta - U -V \right) \delta\left(y + U -V\right)\,.
\eeqa
We denote the Heaviside function by $\theta(x)$ and $\Phi(x,\alpha, \beta)$ is the cdf of the normal distribution $\mathcal{N}(\alpha, \beta)$. Computing the density is straightforward. The result is

\beqa
\rho(y)&=& \frac{1}{\sqrt{2\pi\sigma^2_{\textrm{SR}, \textrm{tot}}} \Phi\left(\Theta, \textrm{SR}_t, \sigma_{\textrm{SR}, \textrm{tot}}\right)} \nn \\
&&\times \exp\left(-\tfrac{1}{\sigma^2_{\textrm{SR}, \textrm{tot}}}\left(y- \textrm{SR}_t(1-2f)\right)^2 \right) \nn\\
&& \times\Phi\left(\sqrt{2\alpha} \nu(y), 0 , 1\right)\,,
\eeqa
where
\beqa
&& \nu(y) = \tfrac{1}{2\alpha}\left( \Theta - \textrm{SR}_t + (1-2f)(\textrm{SR}_t(1-2f)-y)\right)\,, \nn\\
&& 2\alpha = 4\sigma^2_{\textrm{SR}, \textrm{tot}} f(1-f)\,. 
\eeqa

\newpage

\end{document}